\documentclass[aps,pre,twocolumn,superscriptaddress,amsmath,amssymb,longbibliography]{revtex4}

\usepackage{graphicx}

\usepackage{dcolumn}

\usepackage{enumitem,kantlipsum}

\usepackage{bm}
\usepackage[normalem]{ulem}
\usepackage{color}
\usepackage{gensymb}

\usepackage{epstopdf}

\usepackage{xcolor}
\usepackage{color}

\definecolor{green}{rgb}{0,0.5,0}

\begin{document}
 
\title{
Noise crosscorrelations can induce instabilities in coupled driven models}
\author{Sudip Mukherjee}\email{sudip.bat@gmail.com, aca.sudip@gmail.com}
\affiliation{Barasat Government College,
10, KNC Road, Gupta Colony, Barasat, Kolkata 700124,
West Bengal, India}

\begin{abstract}

We study the effects of noise cross-correlations on the steady states of driven, nonequilibrium systems, which are described by two stochastically driven dynamical variables, in one dimension. We use a well-known stochastically driven coupled model with two dynamical variables, where one of the variables is autonomous being independent of the other, whereas the second one depends explicitly on the former. Introducing cross-correlations of the two noises in the two dynamical equations, we show that depending upon the details of the nonlinear coupling between the dynamical fields, such cross-correlations can induce instabilities in the models, that are otherwise stable in the absence of any cross-correlations. { We argue that this is reminiscent of the roughening transition found in the Kardar-Parisi-Zhang equation in dimensions greater than two.} Phenomenological implications of our results are discussed.
\end{abstract}

\maketitle

\section{Introduction}

Time-dependent statistical descriptions of condensed matter systems are often made in terms of continuum, Langevin equations of the relevant dynamical variables driven by noises~\cite{chaikin}. The noises represent inherent microscopic stochasticity of the dynamics. In equilibrium systems, such stochasticity arises from thermal fluctuations. Conditions of thermal equilibrium, known as the fluctuation-dissipation theorem (FDT)~\cite{chaikin}, ensures that the damping in the system is proportional to the noise variance, which is assumed to be Gaussian-distributed with a zero mean. The proportionality constant in fact gives the temperature. In nonequilibrium systems, there is no FDT, and hence, the damping and the noise variance are independent of each other. As a result, nonequilibrium steady states (NESS) are far more diverse and complex than their equilibrium counterparts. 

Physical descriptions of many natural driven systems involve coupled dynamics of several
degrees of freedom. Prominent examples include driven symmetric
mixture of a miscible binary fluid~\cite{binfluid} and magnetohydrodynamics~\cite{mhd}. Stochastically driven binary fluid equations of the velocity and concentration gradient~\cite{binfluid1} and magnetohydrodynamics (MHD) equations of the velocity and magnetic fields~\cite{abmhd} have been used to study turbulence in these systems. In equilibrium systems, conditions of thermal equilibrium, e.g., in the form an FDT, ensures that the noise statistics have no role to play in determining the thermodynamic properties of the system. For instance, relaxational dynamics both without and with a conservation law for the order parameter~\cite{halpin} refer to the same equal-time, thermal equilibrium properties. In contrast to equilibrium systems, in the absence of any FDT, varying the noise statistics can result into distinctly different NESS in driven systems. { For instance, the Kardar-Parisi-Zhang (KPZ) equation driven by white noises~\cite{kpz,stanley}, and its conserved counterpart (the CKPZ equation) driven by conserved noises~\cite{ckpz} have very different universal properties. } 

Introduction of noise cross-correlations in a stochastically driven coupled model necessarily changes the noise distribution, Whether or not this can lead to a new NESS, is a question of basic importance in nonequilibrium statistical mechanics. In fact, there are examples where non-zero cross-correlations of the two noises in the two dynamical equations are found to affect the scaling properties of the NESS. Presence of such noise cross-correlations in driven systems cannot be ruled out on the basis of any symmetry arguments or physical principles.  Simpler reduced models have been proposed and further studied to explore the role of noise cross-correlations. For instance, noise cross-correlations in a nonconserved relaxational model for the complex scalar field turns out to be generally a relevant perturbation on the equilibrium states of the model near a critical point~\cite{niladri-noise}. Subsequently, by using a coupled Burgers model originally proposed in~\cite{abfrey1}, Refs.~\cite{abfrey1,abfrey2} have  shown that noise cross-correlations can lead to continuously varying scaling exponents in the NESS. { Effects of noise crosscorrelations have been studied in MHD turbulence; see Refs.~\cite{abmhd,adzhem}. These crosscorrelations have also been considered in multi-species directed percolation problems~\cite{tauber-book}. Crosscorrelations between different dynamical variables can be potentially important in the scaling properties of complex Ginzburg-Landau equation~\cite{tauber-lieu}, and even in systems of non-condensed matter origin, e.g., economic systems~\cite{mantegna}.} Nonetheless, general understanding of the effects of noise cross-correlations on the NESS of driven models still remains largely at an early stage.

In this work, we revisit the issue of the effects of noise crosscorrelations on the NESS of coupled driven models. Since we are interested in studying a question of principle, it suffices to work with simple models where explicit calculations can be performed relatively easily, but still nontrivial results can be obtained. To that end, we use a one-dimensional (1D) model~\cite{ertas,barabasi,jkb-mode-coup}, where one of the fields $v(x,t)$ is autonomous, and satisfies the well-known 1D Burgers equation~\cite{fns}.  The second dynamical field $b$ is dynamically influenced by $v$, and follows an equation similar to the well-known passive scalar equation with a compressible velocity field. We investigated the universal spatio-temporal scaling properties. Using a specific structure of the noise crosscorrelations, we show that depending upon the specific model at hand, either it introduces instabilities to the NESS characterised by well-known scaling exponents, or is {\em irrelevant} (in the renormalisation group or RG sense), leaving the long wavelength scaling properties unaffected. In the former case, the ultimate steady state that the instability by the noise cross-correlations lead to cannot be determined from the low order perturbation theory employed here. The remainder of this article is organised as follows. In Section~\ref{model}, we have introduced the model we have used, along with the form of the noise cross-correlations. In Section~\ref{gal-inv} we have discussed
the Galilean invariance of the model. Then in Section~\ref{scaling}, we have discussed the details of the dynamic RG analysis on the model. In Section~\ref{summ}, we summarise and conclude. Some of the intermediate technical details including the one-loop Feynman  diagrams are given in Appendix for interested readers.

\section{Model}\label{model}

Noise crosscorrelations can generically exist in any multi-variable system described by stochastically driven dynamical  equations for the dynamical variables.  Since we are trying to answer questions of principles, we use a simple, purpose-built 1D model that suffices for our purposes. The model consists of two dynamical variables $v(x,t)$ and $b(x,t)$. Out of these two, we assume $v$ to be {\em autonomous}, i.e., independent of $b$, and follows the well-known 1D Burgers equation~\cite{fns}. This is given by
\begin{equation}
\frac{ \partial v}{\partial t}=\nu\partial_{xx} v +\frac{\lambda_1}{2}\partial_x v^2 + f_v.\label{eq1}
\end{equation}
Here, $\nu>0$ is a diffusivity, and $\lambda_1$ a non-linear coupling constant, and can be of any sign. Further, $f_v$ is a conserved noises. Hereafter, we refer to $v$ as the ``Burgers velocity field''~\cite{fns}. { The Burgers equation has the same symmetry (Galilean invariance; see below) as the Navier-Stokes equation~\cite{landau-fluid} for viscous fluids, and hence serves as a 1D toy model for ``pressureless fluids''. However, physical representations of the Burgers equation is far wider than that. For instance, it serves as a model for growing nonequilibrium surfaces without overhangs via its mapping to the KPZ equation~\cite{kpz}, the phase in a driven phase-ordered systems with XY-like symmetry~\cite{john-prx}, the longitudinal fluctuations in a directed line or a drifting polymer~\cite{ertas}, when the transverse fluctuations are suppressed.}  

The second field $b$ is assumed to be advected by the Burgers velocity $v$ {\em passively}, i.e., it has no effect on the dynamics of $v$. The dynamics of $b$ follows~\cite{jkb-mode-coup}

\begin{equation}
 \frac{ \partial b}{\partial t}=\mu\partial_{xx} b + \lambda_2 \partial_x (vb) + f_b.\label{eq2} 
\end{equation}
Here, $\mu>0$ is a diffusivity, and $\lambda_2$ a non-linear coupling constant, and can be of any sign. Further, $f_b$ is a  conserved noise. Following the nomenclature of Ref.~\cite{abfrey1}, we call $b$ as the ``Burgers magnetic field''. 
Equations~(\ref{eq1}) and (\ref{eq2}) in fact can be obtained from the coupled Burgers equations without the feedback term in the equation of $v$; see Refs.~\cite{diamond,abfrey1}. These can also be written in terms of nonconserved height or displacement fields; see, e.g., Refs.~\cite{ertas,lahiri}. { The $b$-field being a conserved density may be interpreted as the density of a collection of diffusing particles in contact with or being ``advected by'' $v$. This representation of the model equations (\ref{eq1}) and (\ref{eq2}) actually brings to the class of problems of passive sliders on fluctuating surfaces~\cite{tapas}, which are modeled by the KPZ or the linear Edward-Wilkinson (EW) equation for surfaces~\cite{stanley}.  In these models passive sliders move along the surfaces following particular dynamics rules, often classified as ``advection'' and ``anti-advection''~\cite{tapas}, which in turn should be equivalent to the two signs of the product $\lambda_1\lambda_2$.}

The noises $f_v$ and $f_b$ have zero mean, and are Gaussian-distributed. Their joint probability distribution is characterised by the three variances (written in the Fourier space and functions of frequency $\omega$ and wavevector $k$)
\begin{eqnarray}
 \langle f_v(k,\omega) f_v (-k,-\omega)\rangle &=& 2D_v k^2,\label{var1}\\
 \langle f_b(k,\omega) f_b (-k,-\omega)\rangle &=& 2D_b k^2,\label{var2}\\
 \langle f_v(k,\omega) f_b(-k,-\omega)\rangle &=& 2iD_\times k |k|,\label{var3}
\end{eqnarray}
where $D_v,\,D_b>0$ necessarily, but the sign of $D_\times$ is arbitrary. {  Reality of the noises means the noise variance matrix constructed from (\ref{var1})-(\ref{var3}) must have real, non-negative eigenvalues, which in turn implies $D_v\,D_b\geq D_\times^2$.}
Equation~(\ref{var3}) gives the noise cross-correlation here, which is {\em purely imaginary} and {\em odd} in $k$, i.e., $D_\times(k) = - D_\times(-k)$. The structure of (\ref{var3}) is dictated by the symmetry properties of (\ref{eq1}) and (\ref{eq2}). In line with \cite{abfrey1}, we assume $v$ to be a {\em pseudo-scalar} field and $b$ to be a {\em scalar} field (i.e., a vector and a pseudo-vector at dimensions $d>1$). This means Eqs.~(\ref{eq1}) and (\ref{eq2}) are invariant under $x\rightarrow -x,\,v\rightarrow -v,\,b\rightarrow b$. These symmetries hold true even if there is a ``mean $b$'', i.e., $\langle b\rangle\neq 0$~\cite{abfrey1}. On the other hand, if $\langle b\rangle =0$, the model has a higher symmetry: Equations~(\ref{eq1}) and (\ref{eq2}) are also invariant under $x\rightarrow -x,\,v\rightarrow -v,\,b\rightarrow -b$. While our subsequent calculations specialise for $\langle b\rangle = 0$, we continue to impose invariance under (\ref{eq1}) and (\ref{eq2}) are invariant under $x\rightarrow -x,\,v\rightarrow -v,\,b\rightarrow b$ only. This symmetry implies the cross-correlation function $\langle v(x,t)b(0,0)\rangle$ is an {\em odd} function of $x$. This in turn means, as in Ref.~\cite{abfrey1}, $\langle v(k,t)b(-k,0)\rangle$ is purely imaginary and odd in $k$. Since $\langle v(k,t)b(-k,0)\rangle$ proportional to the noise cross-correlations, and the noises in (\ref{eq1}) and (\ref{eq2}) must follow the symmetries of the corresponding dynamical variables,  (\ref{var3}) follows directly~\cite{foot1}.

\section{Galilean invariance}\label{gal-inv}

Model equations (\ref{eq1}) and (\ref{eq2}) are invariant under a pseudo-Galilean transformation $x\rightarrow x+ v_0t,\, v\rightarrow v+v_0,\, t\rightarrow t,\, \frac{\partial}{\partial t} \rightarrow \frac{\partial}{\partial t} + \lambda_1 v_0\partial_x$, when $\lambda_1=\lambda_2$. When $\lambda_1\neq \lambda_2$, there is no Galilean invariance. However, as our calculations below show (see also Refs.~\cite{ertas,abfrey1,abfrey2,abfrey3} for related discussions) even if $\lambda_1\neq \lambda_2$, i.e., even when they are unequal microscopically, Galilean invariance is recovered and appears as an {\em emergent symmetry} in the long wavelength limit, i.e., $\lambda_1=\lambda_2$ in the renormalised theory, so long as  $\lambda_1\lambda_2>0$ holds. In contrast, if $\lambda_1\lambda_2<0$, Galilean invariance is {\em not} restored even in the long wavelength limit. Hence, Galilean invariance is genuinely broken even in the renormalised theory in this case. These are the two distinct cases, which we discuss separately below. We further consider a third case, in which we set $\lambda_1=0$. Thus in this case, $v$ satisfies the linear 1D diffusion equation. The corresponding equation in terms of a height field is the well-known Edward-Wilkinson equation~\cite{stanley}, forced by a nonconserved noise.

\section{Scaling}\label{scaling}

We are interested in calculating the scaling exponents which characterise the time-dependent correlation functions of $v$ and $b$:
\begin{eqnarray}
 C_v( r,t)\equiv \langle v({ r},t)v(0,0)\rangle&=&|r|^{2\chi_v}f_v(|r|^{z_v}/t),\label{cvreal}\\
 C_b( r,t)\equiv \langle b({ r},t)b(0,0)\rangle&=&|r|^{2\chi_b}f_b(|r|^{z_b}/t).\label{cbreal}
\end{eqnarray}
or their Fourier transformed versions
\begin{eqnarray}
 C_v(k,\omega)\equiv \langle |v({ k},\omega)|^2\rangle&=&k^{2\tilde\chi_v}\tilde f_{v}(k^{z_v}/\omega),\label{cvfour}\\
 C_b(k,\omega)\equiv \langle |b({ k},\omega)|^2\rangle&=&k^{2\tilde\chi_b}\tilde f_{b}(k^{z_b}/\omega),\label{cbfour}
\end{eqnarray}
in the long wavelength limit. Here $\chi_v$  and $z_v$  are the roughness and dynamic exponents of $v(x,t)$; $\chi_b$ and $z_b$ are respectively the corresponding roughness and dynamic exponents of $b(x,t)$; $\tilde\chi_v$ ($\tilde\chi_b$) can be connected to $\chi_v$ ($\tilde\chi_b$) by Fourier transform, giving
\begin{equation}
 \tilde\chi_v=1+\chi_v+z_v,\;\;\tilde\chi_b=1+\chi_b+z_b.
\end{equation}
Further, $f_{v,b}(|r|^z/t)$ and $\tilde f_{v,b}(k^z/\omega)$ are dimensionless scaling functions of their respective arguments.  Notice  that we have allowed for two different dynamic exponents. If $z_v=z_b$, then one gets {\em strong} dynamic scaling, else, if $z_v\neq z_b$, weak dynamic scaling ensues~\cite{dibyendu}. 

\subsection{Linear theory}

The linear limit of the model equations is obtained by setting all nonlinear term to zero, i.e., by setting $\lambda_1 = 0 =\lambda_2$. In this limit, all the two point correlations can be calculated exactly. We have
\begin{eqnarray}
 \langle |v(k,\omega)|^2\rangle &=& \frac{2D_vk^2}{\omega^2 + \nu^2 k^4},\label{lin-corr-u}\\
 \langle |b(k,\omega)|^2\rangle &=& \frac{2D_bk^2}{\omega^2 + \mu^2 k^4},\label{lin-corr-b}\\
 \langle v(k,\omega)b(-k,-\omega)\rangle &=& \frac{2ik|k|D_\times}{(-i\omega + \nu k^2)(i\omega +\mu k^2)}.\label{lin-corr-cross}
\end{eqnarray}
These give the exact exponent values $\chi_v=\chi_b=-1/2$, which may be obtained by inverse Fourier transforming the above correlators, and $z_v=z_b=2$, corresponding to strong dynamic scaling in the linear theory. If noise crosscorrelations vanish, the linearised equations actually admit an FDT. {  In fact, if $D_\times=0$, $v$ and $b$ fully decouple at the linear level, and by using FDT one can identify two ``temperatures'' $T_v=D_v/\nu$ and $T_b=D_b/\mu$ in the linear theory, which are in general unequal. But a non-zero noise cross-correlation breaks FDT even at the linear level, making it impossible to identify any temperature-like quantity. } {  Lastly, it is straightforward to show by using that (\ref{lin-corr-u})-(\ref{lin-corr-cross}) that the ratios of the equal-time correlators $\langle v(x,0)v(0,0)\rangle,\,\langle b(x,0)b(0,0)\rangle,\,\langle v(x,0)b(x,0)\rangle$ are all just numbers.}

\subsection{Nonlinear effects}\label{rg-an}

The presence of the nonlinear terms no longer allows enumeration of the exact scaling exponents for (\ref{eq1}) and (\ref{eq2}), unlike in the linear theory. Thus perturbative treatments are necessary. Na\"ive perturbation theory produces diverging corrections to the model parameters. These divergences may be systematically handled within the framework of dynamic RG~\cite{halpin}. 

While the dynamic RG procedure is already well-documented in the literature~\cite{halpin}, we give below a brief
outline of this method for the convenience of the readers. It is useful to first cast the model equations (\ref{eq1}) and (\ref{eq2}) into a dynamic generating functional by introducing dynamic conjugate fields $\tilde v(x,t)$ and $\tilde b(x,t)$; see Refs.~\cite{tauber-book,janssen}, see also Appendix~\ref{short-action} and Appendix~\ref{rg-appen} for some intermediate details. The dynamic generating functional is then averaged over the Gaussian  distribution of the noises $f_v$ and $f_b$ with variances (\ref{var1}), (\ref{var2}) and (\ref{var3}). The momentum shell dynamic RG procedure consists of integrating over the short wavelength Fourier modes of $v(x,t),\,b(x,t),\,\tilde v(x,t)$ and $\tilde b(x,t)$ in the generating functional. This is then followed by  rescaling of lengths
and time. In particular, we follow the standard approach of initially restricting wavevectors to lie in a Brillouin zone: $|q| < \Lambda$, where $\Lambda$ is an ultra-violet cutoff
 of order the inverse of the lattice spacing $a$,
although its precise value is unimportant so far as the scaling in the long wavelength limit is considered. The  fields $v(x,t),\,b(x,t)$ and their dynamic conjugates $\tilde v (x,t),\,\tilde b(x,t)$ are then split into two parts, a high and low wave vector parts $v(x,t)=v^>(x,t) + v^<(x,t),\,b(x,t)=b^>(x,t) + b^<(x,t)$ and $\tilde v(x,t)=\tilde v^>(x,t) + \tilde v^<(x,t),\,\tilde b(x,t)=\tilde b^>(x,t) + \tilde b^<(x,t)$, where $(v,b)^>(x,t)$ and $(\tilde v,\tilde b)^>(x,t)$ are non-zero in the high wavevector range $\Lambda/b< k < \Lambda,\,b>1$, whereas $(v,b)^<(x,t)$ and $(\tilde v,\,\tilde b)^<(x,t)$ are non-zero in the low wavevector range $k< \Lambda/b$. Next, $v^>(x,t),\,b^>(x,t)$ and $\tilde v^> (x,t),\,\tilde b^>(x,t)$ are to be integrated out in the dynamic generating functional. Of course, this integration cannot be done exactly, but is done
perturbatively in the anharmonic couplings in Appendix~\ref{short-action}.
This perturbation theory is usually represented
by Feynman diagrams, with the order of perturbation
theory given by the number of loops in the diagrams that
we calculate; see Appendix~\ref{rg-appen}. Next to this perturbative step, we rescale length by $x=x' \exp(l)$,
in order to restore the UV cutoff back to $\Lambda$. We further rescale time by $t = t'\exp(l\,z)$, whether or not $z$ is the actual dynamic exponent will be found as we go along. This is then followed by rescaling of $v^<(x,t),\,b^<(x,t)$ and $\tilde v^<(x,t),\,\tilde b^<(x,t)$,
the long wave length parts of $v(x,t),\,b(x,t)$ and $\tilde v(x,t),\,\tilde b(x,t)$; see Appendix~\ref{resc}. We discuss (i) $\lambda_1\lambda_2>0$ (Case I),  (ii) $\lambda_1\lambda_2<0$ (Case II), and (iii) $\lambda_1=0\,\lambda_2> 0$ (Case III) separately below. { These choices defining the three cases have definite physical interpretations in terms of the different classes of dynamics of passive scalars sliding on KPZ or EW surfaces~\cite{tapas}.}

Before we discuss the RG results in details, we note that the dynamics of $v$, independent of $b$, follows the well-known 1D Burgers equation~\cite{fns}, for the scaling exponents $\chi_v$ and $z_v$ are known {\em exactly}, thanks to the Galilean invariance and FDT~\cite{fns,stanley}. This gives $\chi_v=-1/2,\,z_v=3/2$, which are exact. The corresponding scaling exponents of $b$, however, cannot be obtained exactly, necessitating perturbative approaches. The relevant one-loop Feynman diagrams for the model parameters and the noise strengths are given in Appendix~\ref{diag}. Independent of the sign of $\lambda_1,\lambda_2$, the critical dimension of the model is two. Since we are interested in 1D, less than the critical dimension, we use a fixed dimension RG scheme, same as that used in the RG calculations on the KPZ equation to obtain the scaling exponents in 1D~\cite{kpz,stanley}; see also Ref.~\cite{comm1}.

\subsection{Case I: Renormalisation group analysis}

The one-loop Feynman diagrams upon evaluation produces the discrete recursion relations. These are given in Appendix~\ref{dist-rec}. This procedure is followed by rescaling of space, time and the fields together with ${\mathcal L}=e^{dl}\approx 1+dl$ for small $dl$ (here $\mathcal L$ is a running scale factor, not to be confused with the system size, which we formally take to be infinity), which ultimately give the following RG recursion relations:
\begin{eqnarray}
 \frac{d\nu}{dl}&=&\nu\left[z-2+\frac{g}{4}\right],\label{nuflow}\\
 \frac{d\mu}{dl}&=&\mu\left[z-2+\frac{g\psi^2}{2(1+P)P}+ \frac{g(1-P)\psi^2}{(1+P)^2P}\right],\label{muflow}\\
\frac{dD_v}{dl}&=&D_v\left[z-2\chi_v-3 +\frac{g}{4}\right],\label{flowDv}\\
 \frac{dD_b}{dl}&=&D_b\left[z-2\chi_b-3+\frac{g\psi^2}{P(1+P)}-\frac{4\Gamma \alpha g\psi^2}{(1+P)^3}\right],\nonumber\\\label{flowDb}\\
\frac{d\lambda_1}{dl}&=&\lambda_1\left[\chi_v+z-1\right],\label{flowl1}\\
 \frac{d\lambda_2}{dl}&=&\lambda_2\bigg[\chi_v+z-1-\frac{\psi^2 g}{(1+P)^2}+\frac{\psi g(3+P)}{2(1+P)^2}\nonumber \\&-&\frac{g\psi}{2(1+P)}\bigg],\label{flowl2}\\
 \frac{dD_\times}{dl}&=&D_\times\left[z-\chi_v-\chi_b-3\right].\label{flowDx}
\end{eqnarray}
Here, $g\equiv  \frac{\lambda_1^2D_v}{\nu^3}$ is the dimensionless coupling constant, $P\equiv \frac{\mu}{\nu}$ is the dimensionless magnetic Prandtl number, $\Gamma\equiv D_\times^2/D_v^2,\,\alpha\equiv D_v/D_b$. All these are non-negative by construction. {  For reasons of notational convenience, we define $\Phi\equiv \alpha\Gamma$.} In addition, $\psi\equiv \lambda_2/\lambda_1$, which in the present case is positive. {Notice that there are no relevant one-loop corrections to $D_\times$. This is because the vertices in Eqs.~(\ref{eq1}) and (\ref{eq2}) are ${\cal O}(k)$, and hence cannot generate any relevant corrections to the noise crosscorrelation (\ref{var3}), which scales as $k|k|$~\cite{abfrey2}. This reason for  the lack of renormalisation of $D_\times$ actually holds to all orders in the perturbation theory, making $D_\times$ unrenormalised to any order in the perturbation theory.}

Flow equations~(\ref{nuflow})-(\ref{flowDx}) can be used to calculate the flow equations for $g,P,\Gamma,\alpha$. We find
\begin{eqnarray}
 &&\frac{dg}{dl}=g \left[1-\frac{g}{2}\right],\label{flowg}\\
 &&\frac{dP}{dl}=Pg\left[\frac{\psi^2}{2P(1+P)}+\frac{\psi^2(1-P)}{P(1+P)^2}-\frac{1}{4}\right],\label{flowP}\\
 &&\frac{d\psi}{dl}=\psi g \bigg[-\frac{\psi^2 }{(1+P)^2}+\frac{\psi }{(1+P)^2}\bigg], \label{flowpsi}\\
 &&\frac{d\Phi}{dl} = \Phi g\left[-\frac{1}{4}-\frac{\psi^2}{1+P}+\frac{4\Phi\psi^2}{(1+P)^3}\right].\label{flowalgam}
 \end{eqnarray}
 At the RG fixed point, $dg/dl=0=dP/dl=d\psi/dl=d\Phi/dl$. This gives $g^*=2,\,\psi^*=1$ and $P^*=1$ as the {\em stable} RG fixed point; here and below a superscript $^*$ refers to the RG fixed point value of any quantity. {  Notice that $\psi^*=1$ is obtained from (\ref{flowpsi}) for any $P$. Notice that $\psi=0=g$ is {\em also} a fixed point of (\ref{flowg}) and (\ref{flowpsi}); it is however globally unstable.} Unsurprisingly, $g^*$ is same as that for the 1D Burgers equation~\cite{fns,stanley}, since $g$ depends only on the parameters of (\ref{eq1}), which is autonomous. This in turn gives $\chi_v=-1/2,\,z_v=3/2$, which are {\em exact}, due to the Galilean invariance and FDT of the 1D Burgers equation~\cite{fns,stanley,frey-2-loop}; see also Ref.~\cite{sun} for related technical details. Further, $P^*=1$ means at the RG fixed point $\nu^*=\mu^*$, i.e., the two renormalised diffusivities are equal, even if they were unequal microscopically. This further implies that the fields $v$ and $b$ have the {\em same} dynamic exponent: $z_v=z_b=z$. This is an example of {\em strong dynamic scaling}~\cite{dibyendu}. Furthermore, at the RG fixed point $\psi^*=1$ implies that $\lambda_1^*=\lambda_2^*$ at the RG fixed point, such that the fixed point is Galilean invariant, even if it were not so microscopically (but with $\lambda_1\lambda_2>0$ microscopically). Thus, the Galilean invariance is an emergent symmetry, even though it is absent microscopically, a statement that holds even in the presence of noise crosscorrelations just as it does in its absence~\cite{ertas}.  
 
 The RG flow of $\Phi$ requires careful attention. We see that the fixed point condition $d\Phi/dl=0$  produces the following fixed points, at each of which the spatial scaling of $b$ given by $\chi_b$ is analysed:
 
 (i) $\Phi^*=0$, a stable fixed point. At this RG fixed point, the noise crosscorrelations effectively vanish, and the long wavelength scaling properties of the system are identical to that without it: {  $\chi_b=-1/4$}.
 
 (ii) There is a second fixed point  given by the condition
 \begin{equation}
  -\frac{1}{4}-\frac{\psi^2}{1+P}+ \frac{4\Phi\psi^2}{(1+P)^3}=0,
 \end{equation}
giving
   $\Phi^*=3/2\equiv \Phi_{c1}$ obtained by using $\psi^*=1=P^*$, which is {\em linearly unstable}. This implies if $\Phi(\ell=0)<\Phi_{c1}$, the system flows to the fixed point $\Phi^*=0$, meaning  a steady state that is statistically identical to a state having no crosscorrelations at the microscopic level in the long wavelength limit, and hence the scaling exponents have the values identical to their values {\em without} crosscorrelations. {  At this unstable fixed point $\chi_b=-7/8$, different from its value when $\Phi^*=0$, i.e., without crosscorrelations.} However, $\Phi(\ell =0)>\Phi_{c1}$, then the system reaches a steady state not known from perturbation theories. Thus in this case, noise crosscorrelations remain {\em relevant} in the RG sense.
   

 (iii) Given that $\Phi^*=\Phi_{c1}$ is a linearly unstable fixed point,  such that if the ``initial value'' $\Phi(l=0)>\Phi_{c1}$, $\Phi(l)$ flows away as $l$ grows, there should be another ``strong coupling'' fixed point presumably stable, but cannot be accessed in this one-loop perturbation theory. This indicates an instability of the zero crosscorrelation state, induced by noise crosscorrelations. It is instructive to find out how $\Phi(l)$ diverges for sufficiently large $l$. As $\Phi(l)$ grows, (\ref{flowalgam}) reduces to
 \begin{equation}
  \frac{d\Phi}{dl}\approx \frac{4g\Phi^2\psi^2}{(1+P)^3}={\Phi^2}.
 \end{equation}
Solving this, we find
\begin{equation}
 \Phi(l)=\frac{\Phi(\ell=0)}{\ell\Phi(l=0)-1}.
\end{equation}
This shows that $\Phi(\ell)$ diverges as $\ell\rightarrow l_c\equiv 1/\Phi(\ell=0)$ from below. In other words, $\phi(\ell)$ diverges as the system size reaches a {\em non-universal} threshold $a_0\exp(l_c)$, where $a_0$ is a small-scale cutoff. This happens so long as the ``initial'' or microscopic value $\Phi(l=0)>\Phi_{c1}$. What is the nature of the steady state in this case? We note from (\ref{flowDb}) that as $\Phi(\ell)$ grows, $\chi_b$ decreases continuously. This is of course unphysical. We cannot in fact follow the flow of $\Phi(\ell)$ all the way to $\ell\rightarrow l_c$, as the perturbation theory breaks down long before that. 

\begin{figure}[htb]
  \includegraphics[width=7cm]{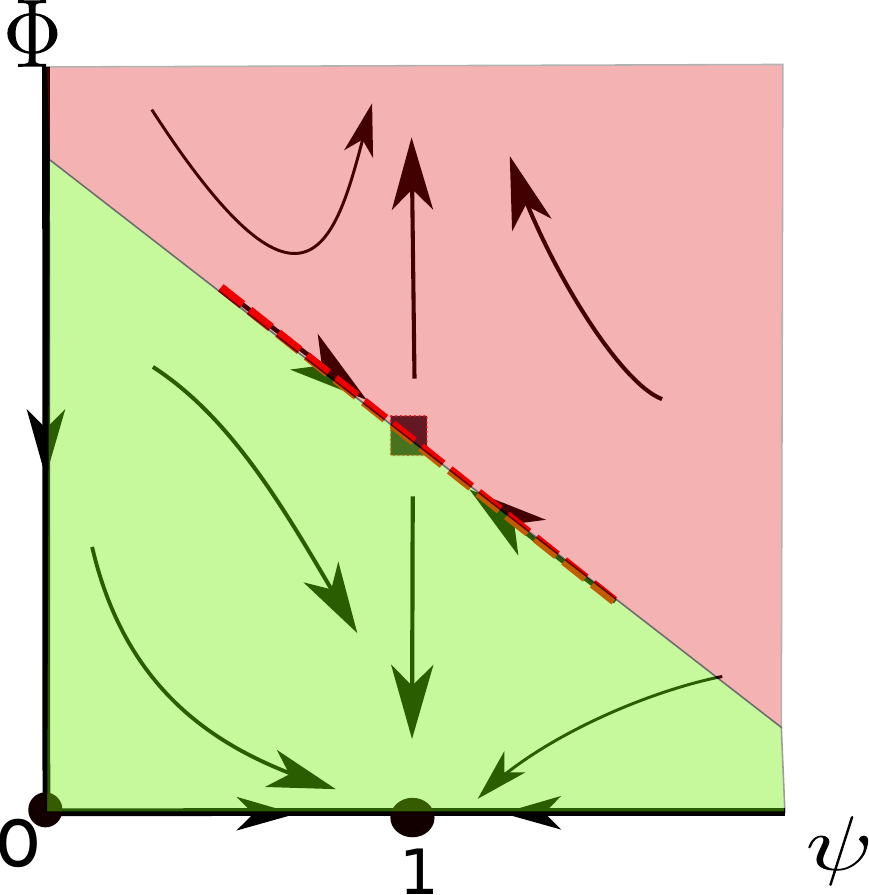}
  \caption{Schematic RG flow diagram in the $\psi-\Phi$-plane for Case I. Arrows show the RG flow directions. The filled square represents the unstable fixed point, and the filled circle on the $\psi$-axis is a stable fixed point. The red broken line is the separatrix  (\ref{sep1}).
  }\label{phase1}
 \end{figure}

{  Combining the fixed points of $\psi$ and $\Phi$, we thus find the following fixed points in the $\psi-\phi$ plane for $g=2$: \\

(i) The origin (0,0). This  has quite an interesting stability property - it is {\em marginally unstable} along the $\psi$-direction, but {\em stable} along the $\Phi$-direction! Naturally, the flow along the $\Phi$-axis is towards the origin.\\

(ii) A globally stable fixed point $(1,0)$. At this fixed point, the long wavelength scaling properties are identical to those with zero noise crosscorrelations.\\

(iii) A fixed point $(1,3/2)$ that is stable along the $\psi$-direction, but unstable along the $\Phi$-direction. \\

(iv) A putative globally stable ``strong coupling'' fixed point, which cannot be captured in our perturbation theory.

The presence of the several fixed points suggests the existence of one or more separatrix, which separates different regions of the phase space having distinct behaviours. Since the origin is stable along the $\Phi$-axis, but unstable along the $\psi$-axis, there should be a separatrix originating from the origin delineating these behaviours. Linearising 
(\ref{flowpsi}) and (\ref{flowalgam}) about (0,0), and defining $\Gamma_1=\Phi/\psi$ as the slope of the separatrix near the origin, we set $d\Gamma_1/dl=0$ for  the separatrix. This gives  $\Gamma_1=0$ near the origin, i.e., the $\psi$-axis. In the same way, we linearise about the unstable fixed point $(1,3/2)$, and define $\tilde \Gamma_1 =\delta \Phi/\delta \psi$, where $\delta \Phi$ and $\delta \psi$ are (small) fluctuations of $\Phi$ and $\psi$ from their fixed point values. For a separatrix that passes through $(1,3/2)$, we set $d\tilde\Gamma/dl=0$ giving 
\begin{equation}
 \delta \Phi = -\frac{3}{4}\delta \psi, \label{sep1}
\end{equation}
giving the separatrix near the fixed point (1,3/2). A schematic RG flow diagram of the model in the $\psi-\Phi$ plane is shown in Fig.~\ref{phase1}. } 

 {  At this point, it is instructive to draw a formal analogy of this instability with the roughening transition in the KPZ equation at $d>2$, a transition between the smooth phase and the perturbatively inaccessible rough phase. This can be reached, e.g., by increasing the noise strength. Similarly in the present problem, by increasing $\Phi$ one can observe a transition from a perturbatively accessible phase having no effects of the noise crosscorrelations to a perturbatively inaccessible phase, where noise crosscorrelations should be relevant in the RG sense, via an unstable fixed point that is reminiscent of a critical point. { Surprisingly, this perturbatively inaccessible phase is expected to be {\em smoother} than the phase without crosscorrelations; see discussions later.} 

 {  While one-loop perturbative RG cannot predict the nature of the steady states near the putative strong coupling fixed point, we note that in the special limit with $\lambda_1=\lambda_2$ and $\nu=\mu$ ``initially'' (i.e., microscopically), these conditions remain satisfied under mode eliminations. It is therefore reasonable to expect that even at the strong coupling fixed point, these should hold in the long wavelength limit at least when these are satisfied microscopically, { i.e., at the small scales, or when the underlying microscopic dynamics in a discrete version of the model equations correspond to these conditions}. This in turn means $z_v=z_b=3/2$ (strong dynamic scaling) at the strong coupling fixed point. Roughness exponent $\chi_b$ however cannot be estimated in this way. Nonetheless, given that $\chi_b\neq \chi_v$ at the unstable fixed point $\Phi^*=\Phi_{c1}$, we are tempted to speculate that $\chi_b\neq \chi_v$ at the strong coupling fixed point as well. Furthermore, since $\Phi$ should be non-zero, at this strong coupling fixed point, we expect at this fixed point $\chi_b < \chi_b(\Phi=0)$. In addition, the topology of the RG flow lines suggest that at the perturbatively inaccessible strong coupling fixed point $\Phi^*> \Phi_{c1}$, the fixed point value of $\phi$ at the unstable fixed point. If this is the case, then we must have the hierarchy $\chi_b ({\rm strong\,coupling})< \chi_b (\Phi=\Phi_{c1})< \chi_b (\Phi=0)$. This runs in contrast to the KPZ equation at $d>2$, where the strong coupling phase is also the ``rough phase'', being rougher than both the smooth phase and at the roughening transition. Physical implication of what this means is not immediately clear to us. Numerical studies of the equations of motion or mode coupling approaches should help in this regard.
 
 The different scaling exponents obtained in Case I are presented in a tabular form in Table~\ref{tab1} below.
  \begin{table}[h!]
 \begin{center}
\begin{tabular}{|p{2cm} |p{2cm}|p{2cm}|p{2cm}| }
  \hline
  \multicolumn{4}{|c|}{Case I fixed points and scaling exponents ($g=2$)} \\
 \hline
 \hline
 $\psi=0,\,\Phi=0$ & $\psi=1,\,\Phi=0$ & $\psi=1,\,\Phi=3/2$ & $\psi:{\rm not known,}\,\Phi\rightarrow {\rm large}$\\
 Linearly unstable along $\psi$, stable along $\Phi$ & Linearly fully stable & Linearly stable along $\psi$, unstable along $\Phi$ & Strong coupling, presumably fully stable\\
 Linear theory for $b$, $\chi_b=-1/2$ & $\chi_b=-1/4,z_b=3/2$ & $\chi_b=-1,z_b=3/2$ &  $z_b=3/2$ if $P=1$ microscopically, $\chi_b$ not known\\
 \hline\hline
 \end{tabular}
\caption{Fixed points in the $\psi-\Phi$ plane and the associated scaling exponents (with $g=2$) in Case I (see text).}\label{tab1}
\end{center}
   \end{table}

Although we cannot obviously follow the RG flow of $\Phi(l)$, starting from $\Phi(l=0)>\Phi_{c1}$, all
the way to infinity as $\Phi(l)$ appears to diverge as $l\rightarrow l_c$ from below, it is possible to speculate about the nature of the
phases in this region of the parameter space. For this, we are guided
by the fact that  $\psi=1$ and $P=1$ are maintained by the perturbation theory, and hence at the strong coupling fixed point also. Given that the form of the noise crosscorrelations, as given in (\ref{var3}), {\em does not} break the Galilean invariance by itself,  this persuades us to speculate that $\psi^*=1$ (corresponding to Galilean invariance) and $P^*=1$ are stable at the strong coupling fixed point. All that the noise crosscorrelations can do is to generate a nonzero fixed point value of $\Phi$, leading to $\chi_b\neq \chi_v$, { a result already holds true at the perturbatively accessible, unstable fixed point, as found in our perturbative RG calculations}. This suggests the existence of {\em another} stable fixed point located at $\psi=1$ and $\Phi>\Phi_{c1}$. This is an Occam's razor-style argument which allows us to draw the simplest RG flow lines that are one hand physically intuitive and also consistent with the perturbatively obtained flow lines; see Fig.~\ref{occ1}. 
 \begin{figure}[htb]
  \includegraphics[width=8cm,height=6cm]{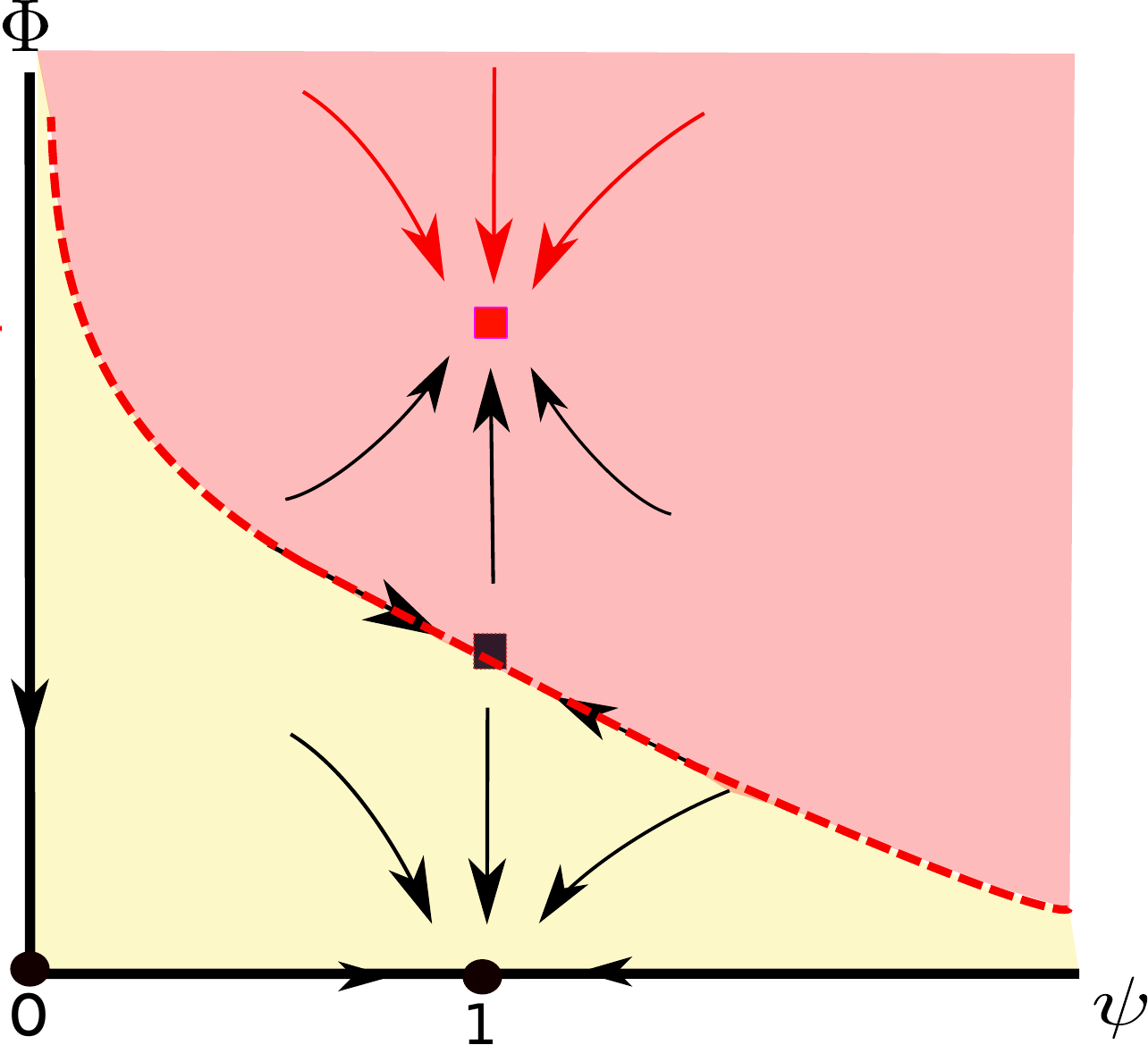}
  \caption{Conjectured global RG flows constructed by using Occam's razor style arguments diagram in the $\psi-\Phi$-plane for Case I. Arrows show the RG flow directions. The filled square represents the unstable fixed point, and the filled circle is a stable fixed point. The red square is the speculated, presumably globally stable fixed point. The red, broken line is the separatrix, drawn schematically as an extension of (\ref{sep1}), which we do not expect to meet any of the axes at any finite distance from the origin. { In particular, it is not expected meet the $\Phi$-axis, as that necessitate a {\em new} fixed point on the $\Phi$-axis (ruled out in the Occam's razor principle). We also do not expect it to connect the origin (0,0), as the linearised flow near the origin discussed earlier does not suggest that.}}\label{occ1}
 \end{figure}
 }

\subsection{Case II: Renormalisation group analysis}

We now consider the case with $\psi <0$, or $\lambda_1\lambda_2<0$. 
{As in Case I, there are no relevant corrections to $D_\times$.} To proceed further, we assume $\lambda_1>0$ without any loss of generality. Then $\lambda_1\lambda_2<0$ implies $\lambda_2<0$. Writing $\psi = -|\psi|$, flow equation (\ref{flowpsi}) takes the form
\begin{eqnarray}
 \frac{d|\psi|}{dl}&=&|\psi|g\bigg[-\frac{\psi^2 }{(1+P)^2}-\frac{|\psi| }{(1+P)^2}\bigg]. \label{flowpsi1}
\end{eqnarray}
Thus, $\psi^*=0$ is the {\em only} RG fixed point, which is {\em stable}. Further, $g^*=2$ at the RG fixed point, which is stable as before. {  With $\psi^*=0$, Eq.~(\ref{eq2}) effectively decouples from Eq.~(\ref{eq1}); as a result, fluctuation corrections to $\mu$ vanish}, which means $z_b=2$. However, the fluctuation corrections to $\nu$ remains unaffected, giving $z_v=3/2$ as before. Therefore $z_v>z_b$, giving $P^*\rightarrow 0$ at the RG fixed point. This gives {\em weak dynamic scaling}~\cite{ertas,barabasi,jkb-mode-coup}. These in turn gives
\begin{equation}
 \frac{d\Phi}{dl}=-\frac{\Phi}{2}<0,
\end{equation}
in the long wavelength limit. This means $\Phi$ flows to zero rapidly near the RG fixed point in the thermodynamic limit. Thus, the effects of the noise cross-corrections are {\em irrelevant} in the RG sense when $\lambda_1\lambda_2<0$: even if noise crosscorrelations are present microscopically, the long wavelength scaling properties of the steady states are same as those without them. A schematically drawn RG flow diagram in the $|\psi|-\Phi$-plane is shown in Fig.~\ref{phase2}. {  Unsurprisingly, the origin (0,0) is the only stable fixed point in the flow diagram. The flows along both the $\psi$ and $\Phi$-directions are towards the origin.} 

\begin{figure}[htb]
  \includegraphics[width=7cm]{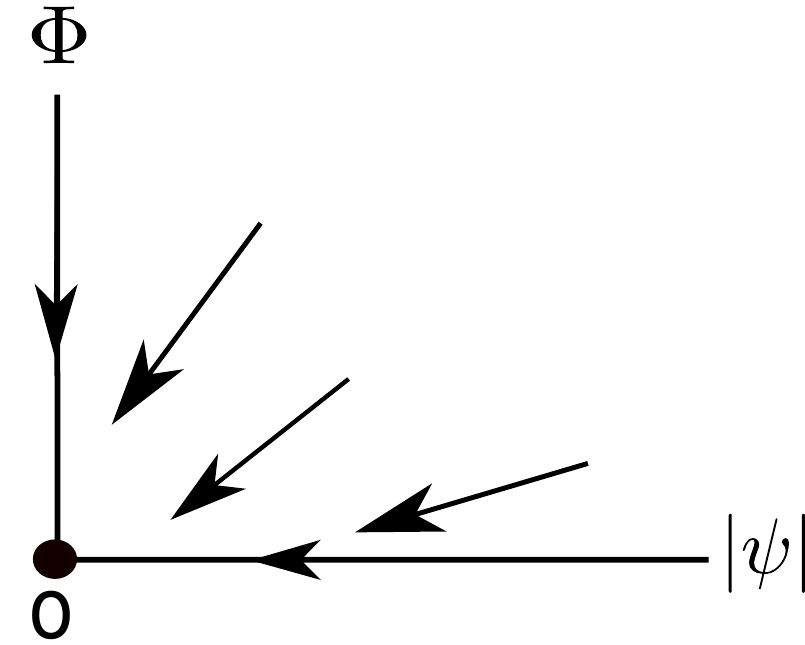}
  \caption{Schematic RG flow diagram in the $\psi-\Phi$-plane for Case II. Arrows show the RG flow directions. The filled circle represents the globally stable fixed point. Here, noise crosscorrelations are generally irrelevant (see text). }\label{phase2}
 \end{figure}

\subsection{Case III: Renormalisation group analysis}

We now consider the case with $\lambda_1=0$. Thus, not only $v$ is autonomous, it follows a linear equation, and its scaling exponents are of course exactly known: we have $\chi_v=-1/2,\,z_v=2$. {  In this case, the sign of $\lambda_2$ has no significance.} We first define a new dimensionless coupling constant $\tilde g=\lambda_2^2 D_v/\nu^3$, which plays the role of $g$ here. The RG flow equations now read 
\begin{eqnarray}
 \frac{d\nu}{dl}&=&\left[z-2\right],\label{nuflow3}\\
 \frac{d\mu}{dl}&=&\mu\left[z-2+\frac{\tilde g}{2(1+P)P}+ \frac{\tilde g(1-P)}{(1+P)^2P}\right],\label{muflow3}\\
\frac{dD_v}{dl}&=&D_u\left[z-2\chi_v-3 \right],\label{flowDv3}\\
 \frac{dD_b}{dl}&=&D_b\left[z-2\chi_b-3+\frac{\tilde g}{P(1+P)}-\frac{4\Phi \tilde g}{(1+P)^3}\right],\nonumber\\\label{flowDb3}\\
 \frac{d\lambda_2}{dl}&=&\lambda_2\left[\chi_v+z-1-\frac{\tilde g}{(1+P)^2}\right].\nonumber\\\label{flowl23}
 \end{eqnarray}
{Since $\lambda_1=0$, there are no diagrammatic corrections to $D_\times$.} Hence, the flow of $D_\times$ follows the same equation (\ref{flowDx}).  We have noted above that $z_v=2$ gives the dynamic exponent of $v$. Does $z_b=2$ for $b$ as well? If it is so, then $P^*$ must be finite at the RG fixed point. The flow of $P$ can be calculated by using (\ref{nuflow3}) and (\ref{muflow3}) as given above. We find
\begin{equation}
 \frac{dP}{dl}=P\tilde g\left[\frac{1}{2P(1+P)}+\frac{1-P}{P(1+P)^2}\right].\label{flowP3}
\end{equation}
This has a stable fixed point at $P^*=3$, meaning $\mu^*=3\nu^*$. This further means that like $v$, even $b$ has a dynamic exponent $z_b=2$. Furthermore, the flow equation for $\tilde g$ reads
\begin{equation}
 \frac{d\tilde g}{dl}=\tilde g\left[1- \frac{2\tilde g}{(1+P)^2}\right].
\end{equation}
Therefore, at the RG fixed point $\tilde g^*=8$, using $P^*=3$. Then, proceeding as for Case I, we find
\begin{equation}
 \frac{d\Phi}{dl}=\Phi\tilde g\left[-\frac{1}{P(1+P)}+\frac{4\Phi}{(1+P)^3}\right]. \label{flowa3}
\end{equation}
Flow equation (\ref{flowa3}) gives the following fixed points, which interestingly are qualitatively similar to Case I: 

(i) We have $\Phi^*=0$, a stable fixed point. At this RG fixed point, noise crosscorrelations are irrelevant in the RG sense, and the long wavelength scaling properties of the model are statistically identical to its zero noise crosscorrelation version~\cite{ertas,barabasi,jkb-mode-coup}. {  In particular, $\chi_b=-1/6$ and $z_b=2$.}

(ii) {  Then there is a linearly unstable fixed point $\Phi^*=4/3\equiv \Phi_{c2}$. Thus, if the initial value $\Phi(l=0)<4/3\equiv \Phi_{c2}$, $\Phi(l)$ flows to zero, rendering noise crosscorrelations irrelevant in the RG sense. At this unstable fixed point, $\chi_b (\Phi_{c2})=-1/2<\chi_b(\Phi=0)$. On the other hand, if  $\Phi(l=0)>\Phi_{c2}$, $\Phi(l)$ grows indefinitely as $l$ grows. Again as in Case I, this indicates an instability, induced by noise crosscorrelations. In fact, proceeding as in Case I, we can show that $\Phi(l)$ diverges as $l\rightarrow \tilde l_c\equiv 2/\Phi (l=0)$, a non-universal value.   

Similar to Case I, a separatrix can be obtained that passes through $(8,4/3)$ in the $\tilde g-\psi$ plane. Following the procedure outlined in Case I, we find
\begin{equation}
 \frac{\delta\Phi}{\delta\tilde g}=0 \label{sep3}
\end{equation}
as the equation for separatrix near the fixed point $(8,4/3)$, where $\delta \tilde g$ and $\delta \psi$ are small deviations of $\tilde g$ and $\psi$ from their fixed point values.}

(iii) Again as in Case I, given that if  $\Phi(l=0)>\Phi_{c2}$, $\Phi(l)$ grows, another (presumably stable) fixed point $\Phi^*>4/3$ should exist, whose actual value cannot be obtained in the present one-loop perturbation theory. Based on our one-loop theory, no inference can be drawn about the scaling properties of this ``strong coupling'' state. However, given that if one has $P=3$ microscopically, it remains so under mode eliminations, we still expect $z_v=z_b=2$. By using arguments similar to Case I, we expect $\chi_b<\chi_b (\Phi=0)$ at this fixed point. By using arguments similar to Case I above, we again expect an analogous hierarchy $\chi_b(\Phi>\Phi_{c2})<\chi_b(\Phi=\Phi_{c2})<\chi_b(\Phi=0)$. Numerical approaches should provide qualitative results and additional physical insight about the strong coupling phase.

{  
The different scaling exponents obtained in Case III are presented in a tabular form in Table~\ref{tab3} below.
  \begin{table}[h!]
 \begin{center}
\begin{tabular}{ |p{2.7cm}|p{2.7cm}|p{2.7cm}| }
  \hline
  \multicolumn{3}{|c|}{Case III fixed points and scaling exponents ($\tilde g=8$)} \\
 \hline
 \hline
 $\Phi^*=0$ & $\Phi^*=4/3$ & $\Phi\rightarrow \infty$\\
  Linearly stable & Linearly unstable & Strong coupling\\
 $\chi_b=-1/6,z_b=3/2$ & $\chi_b=-1/2,z_b=3/2$ &  $z_b=3/2$ if $P=3$ microscopically, $\chi_b$ not known\\
 \hline\hline
 \end{tabular}
\caption{Fixed points and the associated scaling exponents (with $\tilde g=8$) in Case III (see text)}\label{tab3}
\end{center}
   \end{table}
}
   
A schematically drawn RG flow diagram in the $\tilde g-\Phi$-plane is shown in Fig.~\ref{phase3}. {  There is a stable fixed point at $\tilde g=8,\,\Phi=0$ and an unstable fixed point at $\tilde g=8,\,\Phi=4$. Further, it is clear from (\ref{flowa3}) that the $\Phi$-axis (i.e., $\tilde g=0$) is a {\em marginal} direction. Naturally, the origin, another fixed point, is unstable along the $\tilde g$ direction, but marginal along the $\Phi$-direction.}


\begin{figure}[htb]
  \includegraphics[width=7cm]{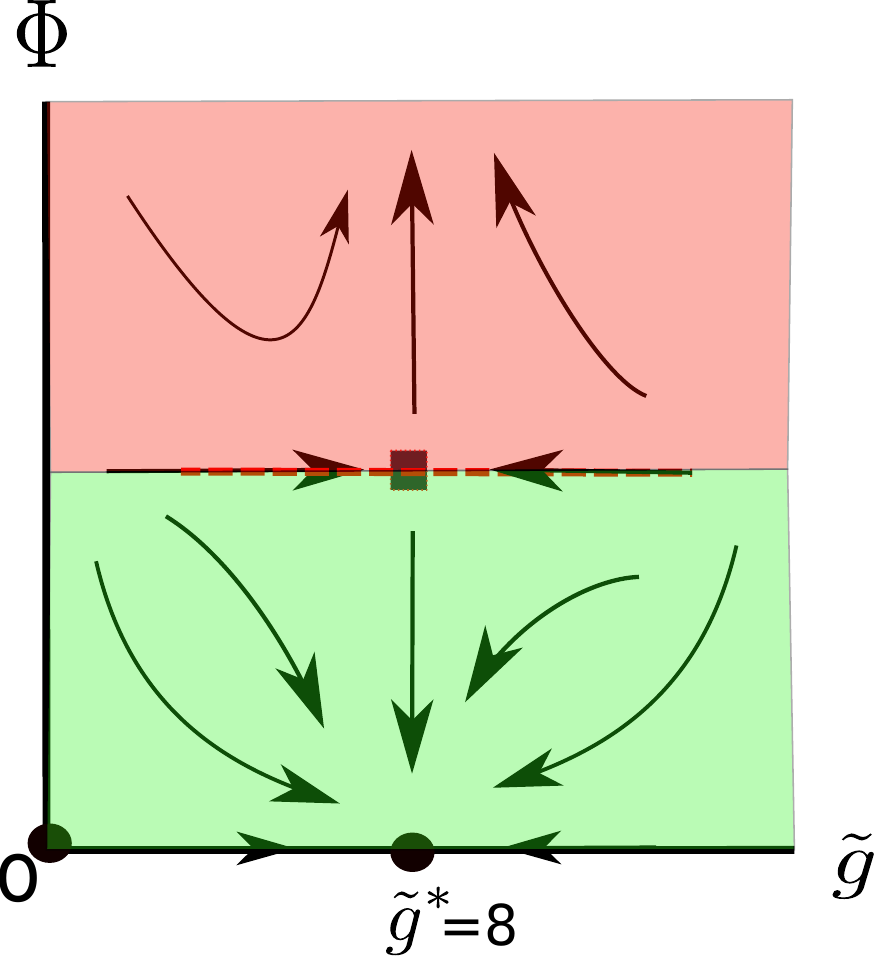}
  \caption{Schematic RG flow diagram in the $\tilde g-\Phi$-plane for Case I. Arrows show the RG flow directions. The filled square represents the unstable fixed point, and the filled circle represents a stable fixed point. The red broken line is the separatrix (\ref{sep3}). }\label{phase3}
 \end{figure}

Notice that in Case III, similar to Case I, the divergence of $\Phi(l)$ for a sufficiently large initial value $\Phi(l=0)$ is reminiscent of the divergence of the dimensionless coupling constant in the higher dimensional ($d> 2$) KPZ equation in the long wavelength limit, resulting into a perturbatively inaccessible strong coupling rough phase. In an analogy to the rough phase in the $d>2$ KPZ equation, we are tempted to speculate the perturbatively inaccessible steady states with a large but presumably finite $\Phi$ as a type of strong coupling  phase.  

{  Similar to Case I, one may use Occam's razor type arguments draw the global RG flow lines, which match with the flow diagram in (\ref{phase3}). Due to the obvious qualitative similarity between  Case I and Case III, such a global RG flow diagram in the $\tilde g-\Phi$ plane in Case III should have the same topology as the corresponding diagram for Case I in its $\psi-\Phi$ plane as shown in Fig.~\ref{occ1}.} 



\section{Summary and outlook}\label{summ}

In this work, we have studied the effects of noise crosscorrelations on the steady states of a 1D coupled driven model. Specifically, one of the dynamical variables $v$ follows the well-known Burgers equation, and evolves autonomously, being independent of the second dynamical field $b$. The second dynamical field $b$ is passively advected by the ``Burgers velocity'' $v$, and follows an equation that closely resembles the well-known passive scalar model~\cite{passive-scal}. We have analysed the long wavelength properties of this model in the presence of finite noise crosscorrelations, whose effects depends upon the precise nature of the model. We consider three different cases, delineated by the nonlinear coupling constants. For instance in Case I with Galilean invariance appearing as an emergent symmetry, where the advective coupling constant $\lambda_2$ in the $b$-equation has the same sign as the advective nonlinearity $\lambda_1$ in the Burgers equation for $v$, a sufficiently strong noise crosscorrelation amplitude above a finite threshold can destabilise the system, whereas its microscopic values weaker than the threshold render noise crosscorrelations irrelevant in the RG sense. In the latter case, the model is identical to the one without noise crosscorrelations in the long wavelength limit. In contrast, in Case II, where $\lambda_2$ and $\lambda_1$ have opposite signs, noise crosscorrelations are {\em generically irrelevant} in the RG sense. We have also considered yet another case denoted Case III  here, where $\lambda_1=0$, making $v$ follow the linear diffusion equation. In this case, similar to Case I, noise crosscorrelations with amplitudes greater than a threshold lead to instabilities, whereas when below the threshold, it is irrelevant in the RG sense.  In the unstable cases, the eventual steady states cannot be obtained from our calculations. Numerical simulations of equivalent lattice models, or direct numerical solutions of the model equations can verify our perturbative results, and also shed light on the instabilities and the resulting unknown steady states. {  The existence of unstable fixed points in Case I and Case III, and the associated perturbatively inaccessible putative strong coupling phases have strong resemblance with the well-known roughening transition in the KPZ equation at $d>2$.  Quite interestingly, the scaling exponents of $b$ at the unstable fixed point in both Case I and Case III are less than their values at zero noise crosscorrelations. This suggests that the field $b$ actually {\em fluctuates less} at the unstable critical point. The RG analysis fails to capture the strong coupling phases in Case I and Case III. Mode coupling methods may be useful in extracting the scaling exponents in these strong coupling phases~\cite{jkb-mode-coup}. } It may however be noted that in Case III there are no parameter regimes where mode coupling theories may be straightforwardly applied to the strong coupling regime. This is because in Case III vertex corrections play a significant role for any nonzero $\lambda_2$. This is in contrast to Case I, where if the model is Galilean invariant (i.e., $\lambda_1=\lambda_2$) microscopically, there are no relevant vertex corrections. {On the whole, we thus see that the precise effects of the noise crosscorrelations depend quite sensitively on the details of the models under consideration.} { Considering the specific case of passive particles sliding on fluctuating surfaces, we note that different degrees of clustering were found for the advection and anti-advection cases when the surface is a KPZ surface, and also when the surface is an EW surface~\cite{tapas}. How noise crosscorrelations can be included in these studies, and what new effects they may bring in are important questions to study. We hope our work will provide new impetus to further studies along this direction.}

In this work, we have confined ourselves to studying only conserved noises. Similar calculations as here can be performed with noise variances that scale differently with $k$, e.g., long-range noises. { If all the noise correlations defined in (\ref{var1})-(\ref{var3}) become singular for small-$k$ (instead of being conserved noises), there will be no renormalisation of the noise variances, giving us exact exponent identities~\cite{medina}. However, the amplitude $D_b$ can still be affected (in particular get reduced) by $ D_\times$ (see, e.g., Ref.~\cite{abmhd} for similar effects). A semblance of an instability may be created if the effective $D_b$ can be turned negative this way. We look forward to further investigations in this direction.} 

{ We note that the physical effects discussed in the present manuscript will not arise in an analogous coupled nonconserved model, as the fluctuations of any nonconserved variables (away from any critical points) are short-lived and hence are non-hydrodynamic variables, rendering them irrelevant (in the RG sense). In this case, the linear theory scaling holds.  Nonetheless, if there are critical points in the nonconserved models, the critical point fluctuations are long-lived and can alter the linear theory scaling behaviour. Whether or not the eventual scaling behaviour will be similar to those in the present studies will depend upon the specific nature of the model system. In this regard, it will be useful to construct appropriate coupled models and study them systematically.}

{ It will be interesting to consider possible $d$-dimensional generalisations of the results presented here. There are however some intrinsic difficulties involved stemming from the lack of analytical knowledge on the rough phase of the Burger's equation at $d\geq 2$~\cite{stanley}. At 2D, the Burger's equation only has a rough phase not accessible to perturbative RG. As a result, how the noise crosscorrelations can affect the dynamics of $b$ is no longer perturbatively calculable due to the coupling of $b$ with $v$. At dimensions $d>2$, the Burger's equation under goes a nonequilibrium roughening transition between a smooth phase that is statistically identical with the Edward-Wilkinson (EW) model and a perturbatively inaccessible rough phase. No conclusion on the effects of the noise crosscorrelations on the $b$-field by means of perturbative RG can be drawn when the Burger's equation is in its rough phase. Instead, mode coupling approaches should be helpful~\cite{jkb-mode-coup}. However, when the Burger's equation is in its smooth phase, the problem essentially reduces to a $d$-dimensional version of Case III (coupling with the EW  model), in which case perturbative RG may be used. These will be studied separately in the future.} 

We have assumed the noise crosscorrelations to be imaginary and odd in Fourier wavevector $k$. What would happen if we chose a different structure for the noise crosscorrelations, e.g., real and even in $k$? Straightforward perturbation theory generates a term of the form $\partial_{xx} v$ in the fluctuation-corrected (\ref{eq2}). This is not surprising, since a real and even noise crosscorrelation implies that $v$ and $b$ have the same properties under parity inversions; {see, e.g., discussions in Ref.~\cite{abfrey1}}. This then no longer rules out a diffusive $v$ term in (\ref{eq2})~\cite{abfrey1}. Given that $v$ is autonomous, such a term in (\ref{eq2}) effectively acts like another additive noise for $b$, whose statistics is given by the statistics of $v$. It would be interesting to study how  noise crosscorrelations in that case affects the known scaling and stability of the NESS without it.

We have used a simple, purpose-built, minimal coupled 1D model to study the role of noise crosscorrelations.   However, the question of the effects of noise crosscorrelations should be important in hosts of natural systems, going much beyond such simple driven 1D models. RG studies, for instance, on  active XY models~\cite{astik1,astik2}, passive scalar models~\cite{sudip-multi}, can give interesting insights on the precise role of noise crosscorrelations on the NESS of more complex models. We hope our work here will provide impetus on further studies on this topic. However, the question of crosscorrelations can arise in many systems of different physical origins. A particularly interesting problem could be where the stochasticity in the dynamics would be crosscorrelated with the background quenched disorder. This may arise, e.g., in quenched disordered driven models; see, e.g, Ref.~\cite{sm-ckpz}.}

{\em Acknowledgement:-} S.M. thanks 
the SERB, DST (India) for partial financial support through the TARE scheme [file no.: TAR/2021/000170] (2021).

\appendix

\section{Action functional}\label{short-action}

We give the action functional ${\cal S}$ corresponding to Eqs.~(\ref{eq1}) and (\ref{eq2}). It is defined via the generating functional $\cal Z$ given by
\begin{equation}
 {\cal Z}=\int {\cal D}v{\cal D}b{\cal D}\tilde v{\cal D}\tilde b \exp (-{\cal S}),
\end{equation}
where $\cal S$ is given by

\begin{widetext}

\begin{eqnarray}
 {\cal S}&=&-\int\frac{dk}{2\pi}\frac{d\omega}{2\pi}\left[D_vk^2 |\tilde v(k,\omega)|^2 + D_bk^2 |\tilde b(k,\omega)|^2 + iD_\times k|k|\tilde v(k,\omega)\tilde b (-k,-\omega)\right]\nonumber \\
 &-& \int dx dt\left[\tilde v\left(\partial_t v - \frac{\lambda_1}{2}\partial_x v^2 - \nu\partial_{xx}v\right) +\tilde b\bigg(\partial_t b -\lambda_2\partial_x(vb)-\mu\partial_{xx}b\bigg)\right].\label{act-func}
\end{eqnarray}

\end{widetext}

\section{Details of the RG analysis}\label{rg-appen}
In this Section, we discuss some details of the momentum shell RG procedure applied on our model.

\subsection{Feynman diagrams}\label{diag}

\begin{figure}[htb]
\includegraphics[width=5cm]{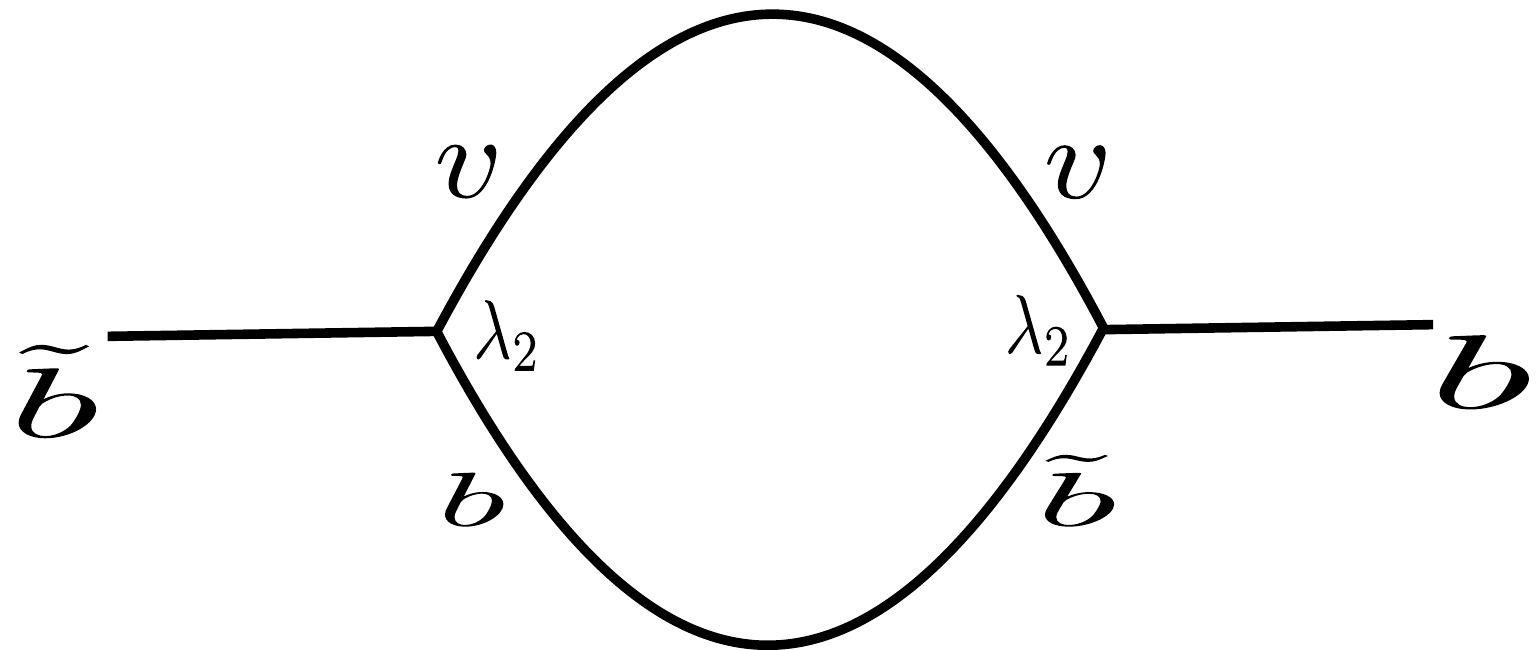}
 \caption{One-loop Feynman diagram that corrects $\mu$.}\label{mu-diag}
\end{figure}

\begin{figure}[htb]
 \includegraphics[width=5cm]{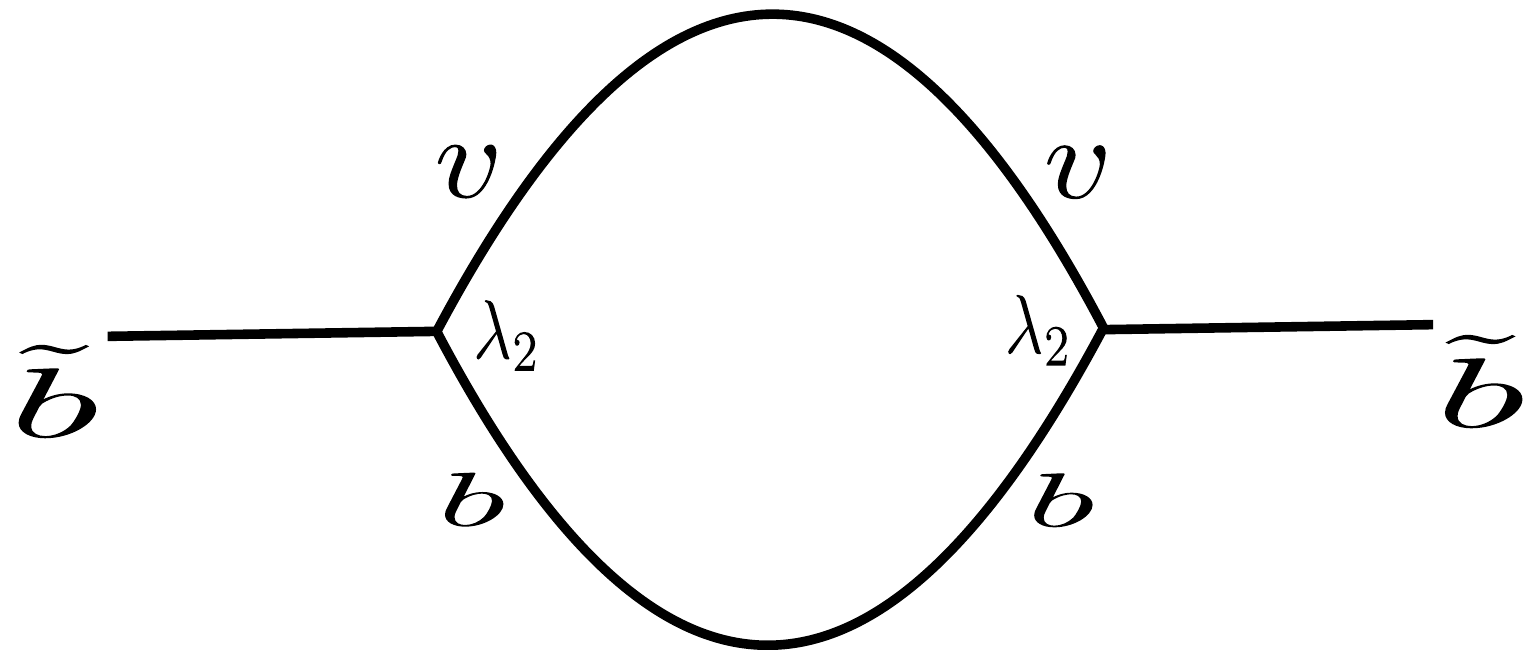}\\
 \includegraphics[width=5cm]{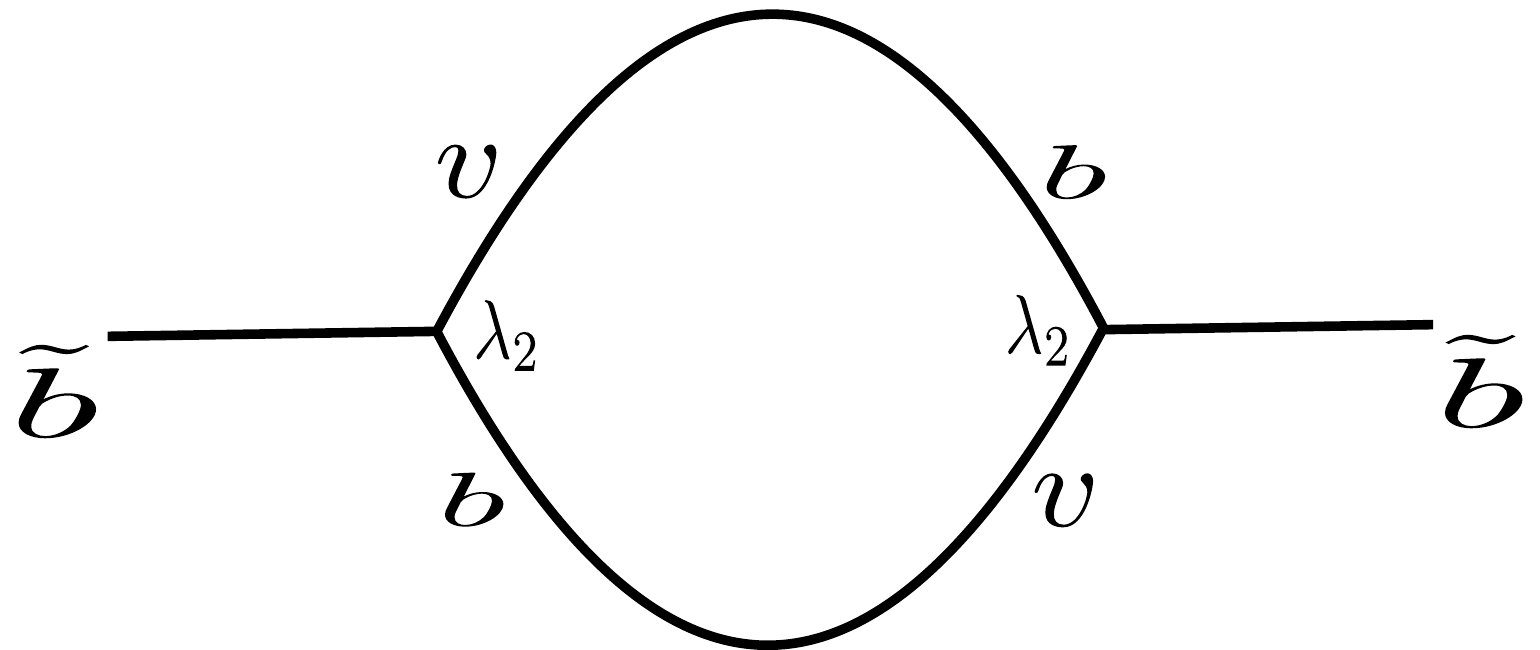}
 \caption{One-loop Feynman diagrams that correct $D_b$. The bottom diagram carries the contribution from the noise crosscorrelations (see text).}\label{Db-diag}
\end{figure}

\begin{figure}[htb]
 \includegraphics[width=5cm]{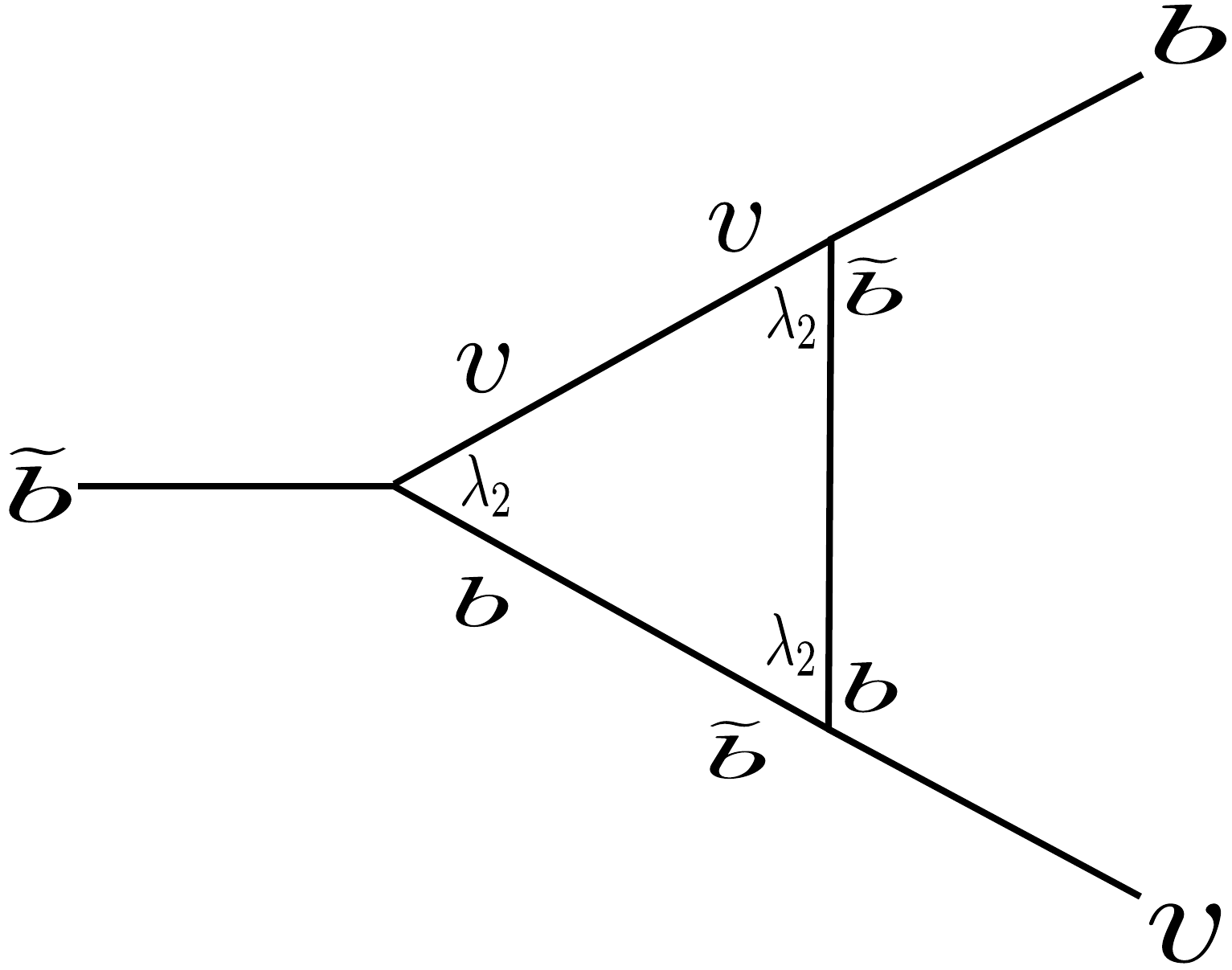}\\
 \includegraphics[width=5cm]{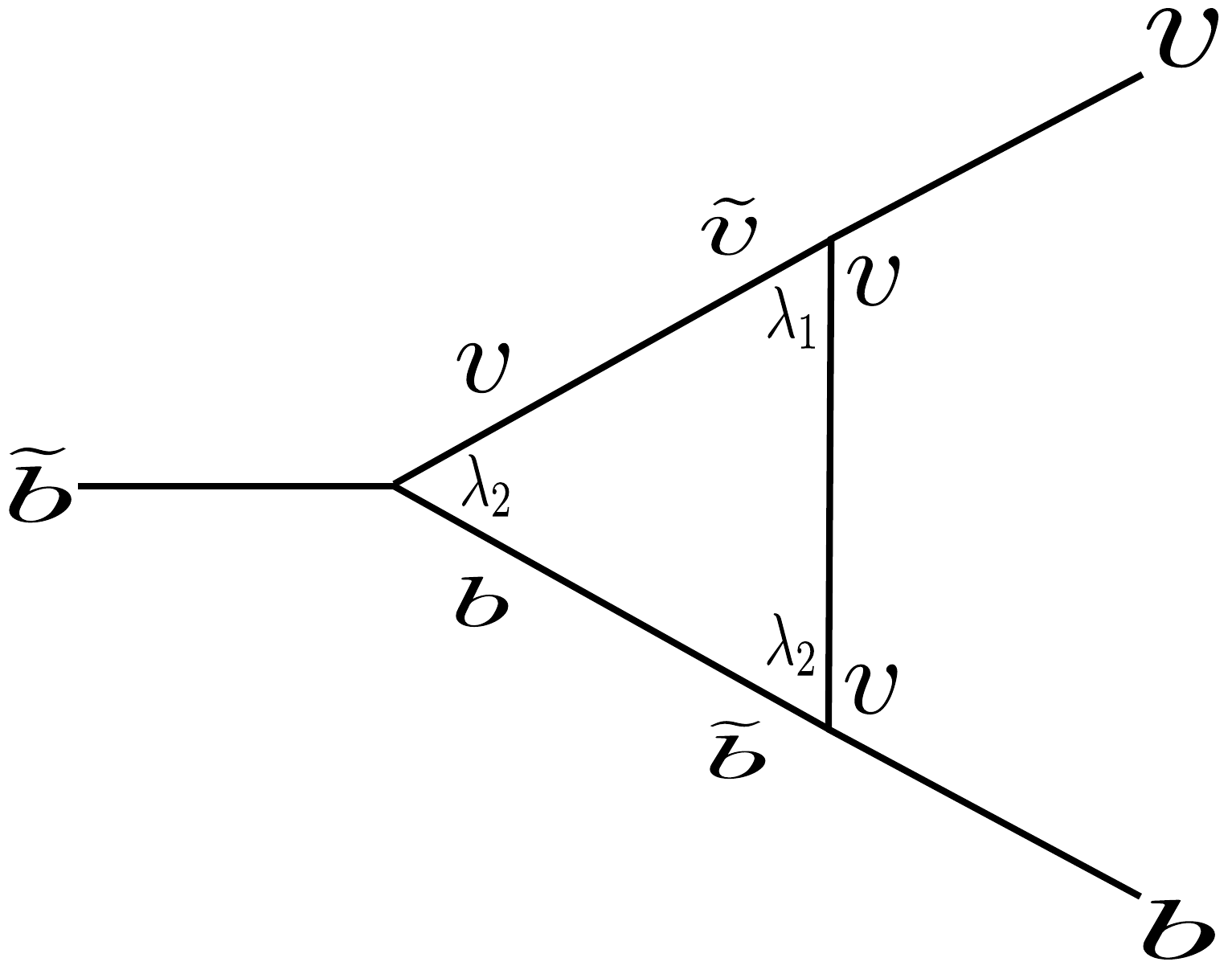}\\
 \includegraphics[width=5cm]{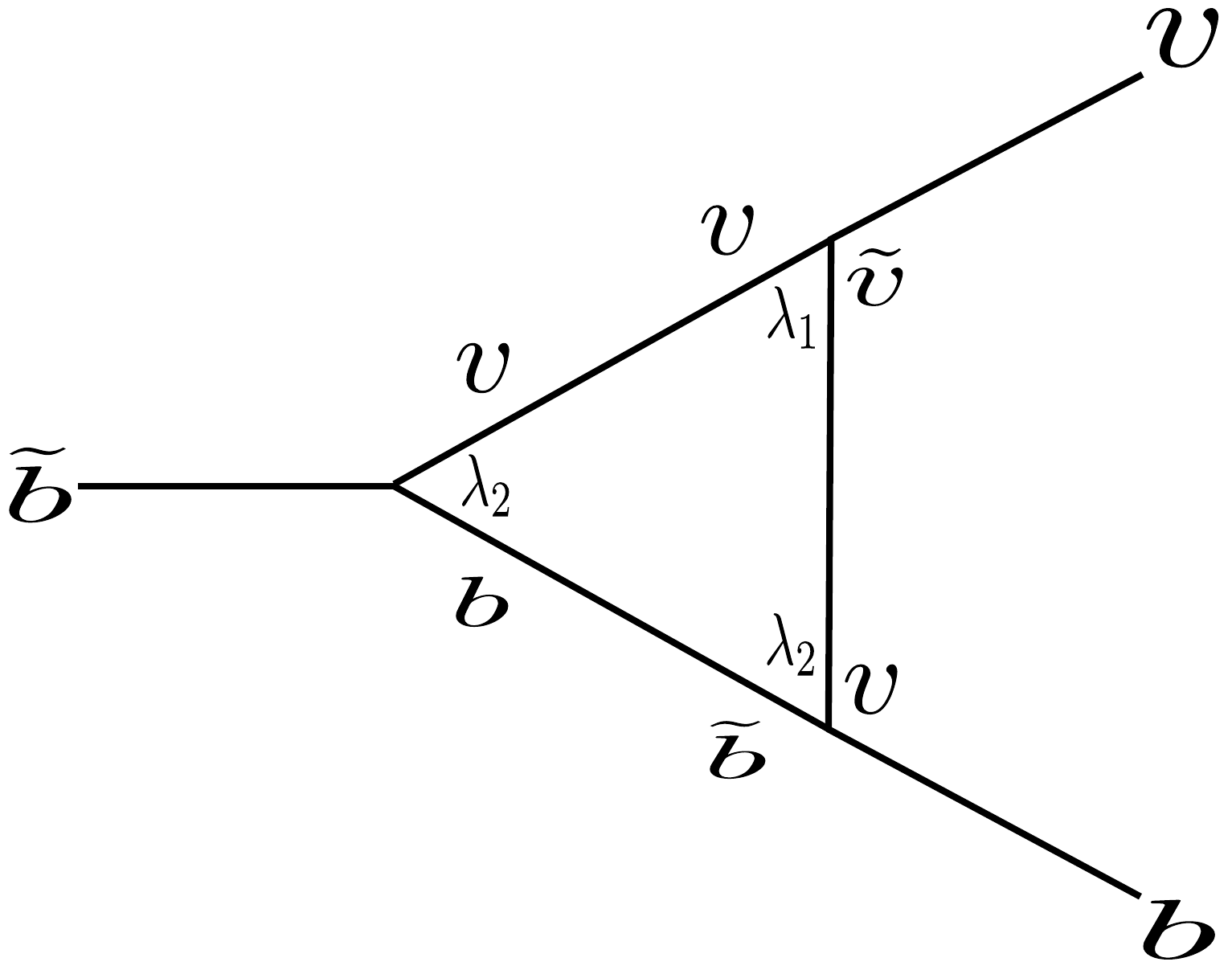}
 \caption{One-loop Feynman diagrams which correct $\lambda_2$.}\label{l2-diag}
\end{figure}

In this Section, we give the one-loop Feynman diagrams for the model parameters in (\ref{eq2}). The corresponding one-loop diagrams are standard; see, e.g., Refs.~\cite{stanley}. We show the one-loop diagrams in Figs.~\ref{mu-diag}, \ref{Db-diag} and \ref{l2-diag}. 

\subsection{Discrete recursion relations}\label{dist-rec}

\begin{eqnarray}
 D_v^< &=& D_v + \frac{\lambda_1^2 D_v^2}{4\nu^3}\frac{\delta l}{\Lambda\pi},\\
 D_b^<&=& D_b+\bigg[\frac{\lambda_2^2 D_vD_b}{\nu\mu(\nu+\mu)} -\frac{4\lambda_2^2 D_\times^2}{(\nu+\mu)^3}\bigg]\frac{\delta l}{\Lambda\pi},\\
 D_\times^< &=& D_\times,\\
 \nu^<&=&\nu+\frac{\lambda_1^2D_v}{4\nu^2}\frac{\delta l}{\Lambda\pi},\\
 \mu^<&=&\mu +\left[\frac{\lambda_2^2D_v}{2\nu(\nu+\mu)}+\frac{\lambda_2^2D_v(\nu-\mu)}{\nu(\nu+\mu)^2}\right] \frac{\delta l}{\Lambda\pi},\\
 \lambda_1^<&=& \lambda_1,\\
 \lambda_2^<&=&\lambda_2 - \bigg[\frac{\lambda_2^3 D_v}{\nu(\nu+\mu)^2} - \frac{\lambda_2^2\lambda_1(3\nu+\mu)D_v}{2\nu^2(\nu +\mu)^2}\nonumber \\ &+&\frac{\lambda_2^2\lambda_1D_v}{2\nu^2(\nu +\mu)}\bigg]\frac{\delta l}{\Lambda\pi}.
\end{eqnarray}

\subsection{Rescaling of space, time and fields}\label{resc}

We perform the following rescaling of space, time and the fields:
\begin{equation}
 x\rightarrow \exp (l)\,x,\,t\rightarrow \exp(l\,z)t,\,v\rightarrow \exp(l\chi_v)v,\,b\rightarrow \exp(l\chi_b) b.
\end{equation}

Under these rescalings, the different model parameters scale as follows: $(\nu,\mu)\rightarrow \exp[(z-2)l] (\nu,\mu),\, D_v\rightarrow \exp[(z-2\chi_v-3)l] D_v,\,D_b\rightarrow \exp[(z-2\chi_b-3)l] D_b,\,D_\times\rightarrow \exp[(z-\chi_v-\chi_b-3)l] D_\times,\,(\lambda_1,\lambda_2)\rightarrow \exp[(z+\chi_v-1)l](\lambda_1,\,\lambda_2)$.

\subsection{Additional technical remarks}

\begin{figure}[htb]
 \includegraphics[width=6cm]{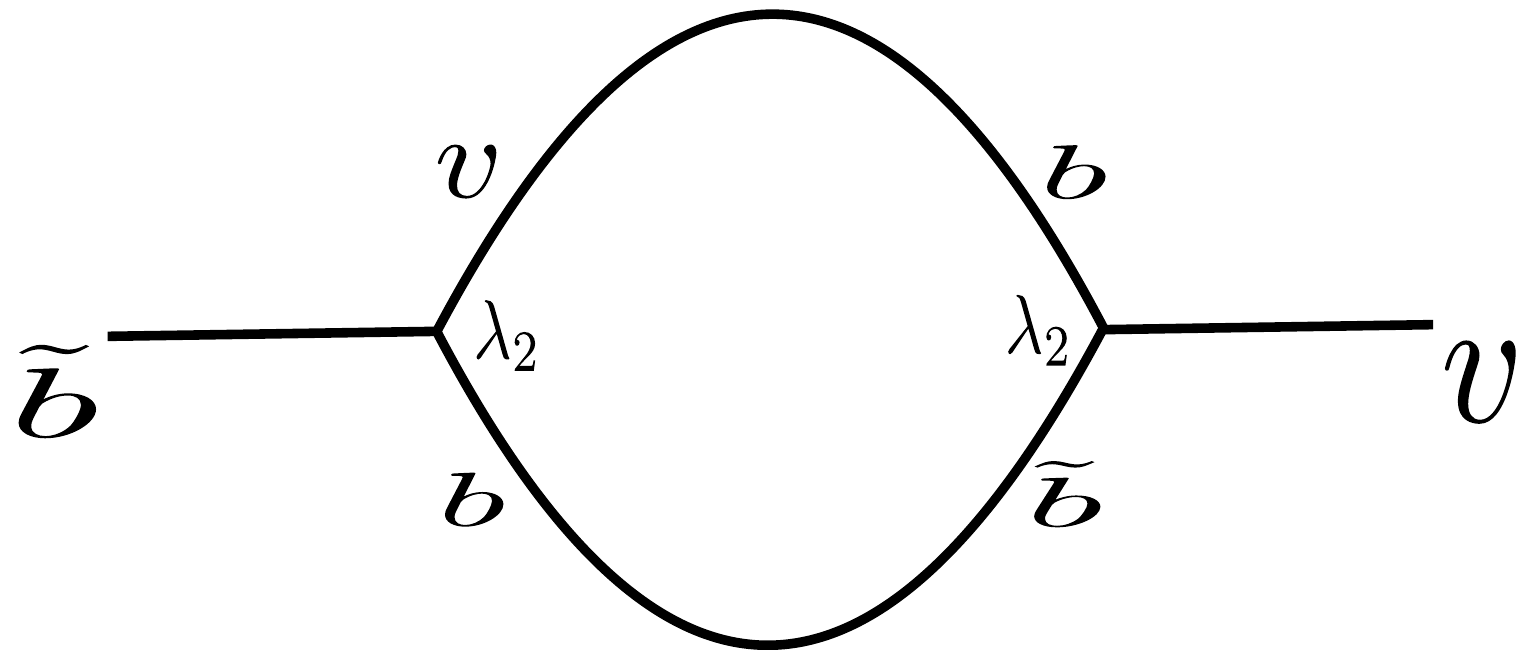}\\
 \includegraphics[width=6cm]{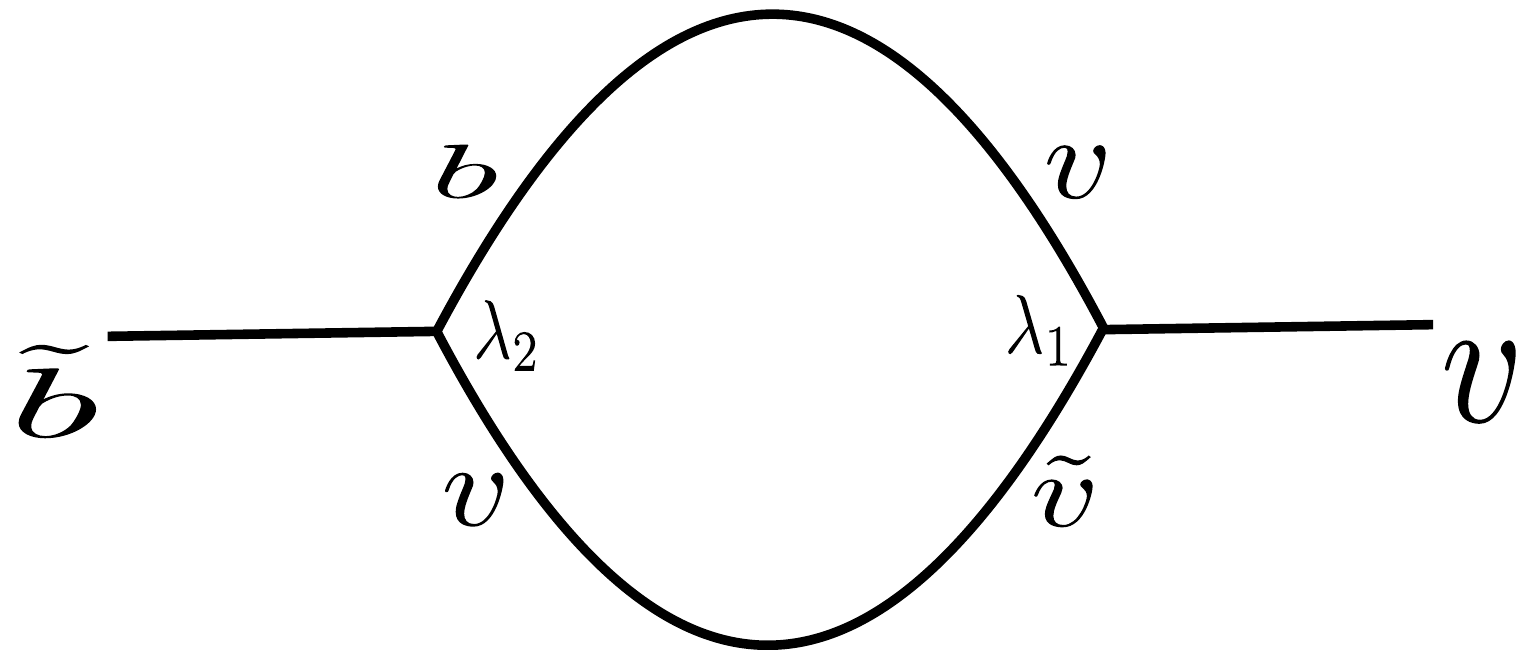}
 \caption{One-loop diagrams apparently generating a mean magnetic field-like term in the $b$ equation (see text).}
 \label{mean-diag}
\end{figure}

We close with a few technical points. Throughout this work, we have assumed $\langle b\rangle =0$. This condition must then be ensured in the successive steps of RG. Since $b$ follows a conservation law, $\langle b(x,t)\rangle$, which is nothing but $\langle b(k=0,t)\rangle$, is a constant of motion.  It however turns out that, if $D_\times\neq 0$, a non-zero $\langle b\rangle$ is actually generated under RG in Eq.~(\ref{eq2}). This is represented by the Feynman diagrams in Fig.~\ref{mean-diag}. Evaluating this, we get
\begin{equation}
 \left[\frac{ikD_\times\lambda_2^2}{\nu(\nu+\mu)}-\frac{ikD_\times\lambda_1\lambda_2}{\mu(\nu+\mu)}\right]\frac{\delta l}{\pi\Lambda},
\end{equation}
which is in general non-zero and contributes to an apparent ``mean magnetic field'' in Eq.~(\ref{eq2}). Including
a counterterm to cancel the spurious $\langle b\rangle$ automatically
ensures the absence of such a term in the theory. For Case I, this contribution vanishes at the RG fixed point; for Case II and Case III, they  do not. Nonetheless, inclusion of an appropriate counterterm ensures that the physical $\langle b\rangle$ remains zero order-by-order~\cite{abfrey1}.

We have also assumed in the course of our calculations that $\langle v\rangle =0$, i.e., there is no ``mean Burgers velocity flow''. A nonzero mean flow would generate linear propagating modes in (\ref{eq1}) and (\ref{eq2}). This is a redundant point in Case I, where at the Galilean invariant fixed point, such mean flow-induced propagating modes can be absorbed by going to a comoving frame. In Case II and Case III, there is no single comoving frame, where both these propagating modes can be absorbed. However, individual propagating modes can be separately absorbed by going to the corresponding comoving frames. Focusing on the comoving frame of $b$, straightforward power counting reveals that the remaining propagating mode in (\ref{eq1}) reduces the infra-red divergences of the one-loop diagrams for the parameters in (\ref{eq2}), which necessarily involves both $v$- and $b$-lines; see, e.g., Ref.~\cite{dibyendu} for related discussions in a coupled driven model. This in turns renders $\lambda_2$ irrelevant in the RG sense, making $b$ autonomous in the long wavelength limit. Naturally, noise crosscorrelations have no visible effects on the steady states of $b$ in this case.


\begin{thebibliography}{99}
\bibitem{foot1} As argued in Ref.~\cite{abfrey1}, invariance under  $x\rightarrow -x,\,v\rightarrow -v,\,b\rightarrow b$ together with a cross-correlation that is real and even in $k$ allows cross dissipative terms, i.e., a $\partial_{xx} v $-term in (\ref{eq2}) and a $\partial_{xx}b$ in (\ref{eq1}) on symmetry ground. Indeed, these are generated in a perturbaion theory.
\end{thebibliography}
\end{document}